\pdfoutput=1
\documentclass[aps,prx,floatfix,nofootinbib,superscriptaddress,twocolumn,10pt,raggedbottom]{revtex4-2}
\usepackage{newpxtext}
\usepackage[utf8]{inputenc}
\usepackage{graphicx,graphics}
\usepackage{dcolumn}
\usepackage{amsmath,amssymb,amsfonts}
\usepackage{amssymb}
\usepackage{amsthm}
\usepackage{latexsym,verbatim}
\usepackage{bm,dsfont}
\usepackage{ulem}
\usepackage{dsfont}
\usepackage{overpic}
\usepackage[dvipsnames]{xcolor}
\usepackage[breaklinks=true,colorlinks,citecolor=Violet,linkcolor=Violet,urlcolor=Violet]{hyperref}
\usepackage{array}
\usepackage{float}
\usepackage{nicefrac}
\usepackage{overpic}
\usepackage{pifont}
\usepackage{tikz}
\usepackage{quantikz}

\newtheorem{definition}{Definition}

\newtheorem{conceptual}{Conceptual step}

\usepackage{tcolorbox}
\tcbset{coltitle=black,fonttitle=\bfseries,
  colback=blue!7!white,colframe=gray!25!white,width=1\columnwidth,valign=center,
  left=1mm, right=1mm, top=1mm, bottom=1mm}

\begin{document}

\setcounter{secnumdepth}{2}
\setlength{\parskip}{0.1pt}

\title{
Global control via quantum actuators
}

\author{Roberto Menta}
\email{roberto.menta@sns.it}
\affiliation{NEST, Scuola Normale Superiore, I-56127 Pisa, Italy}
\author{Francesco Cioni}
\affiliation{NEST, Scuola Normale Superiore, I-56127 Pisa, Italy}
\author{Riccardo Aiudi}
\affiliation{Dipartimento di Fisica dell’Universit\`{a} di Pisa, Largo Bruno Pontecorvo 3, I-56127 Pisa, Italy}
\author{Marco Polini}
\affiliation{Dipartimento di Fisica dell’Universit\`{a} di Pisa, Largo Bruno Pontecorvo 3, I-56127 Pisa, Italy}
\author{Vittorio Giovannetti}
\affiliation{NEST, Scuola Normale Superiore, I-56127 Pisa, Italy}

\begin{abstract}
We introduce the concept of quantum actuators as mediators for globally controlled quantum computation. Auxiliary quantum systems act as controllable elements that transiently store and release interaction energy, enabling the selective activation of multi-qubit gates within globally driven architectures. During compilation they remain passive and require no fine-grained local control, while during operation they allow for controlled activation of interactions and directional flow of quantum information. We provide a framework for embedding quantum actuators in globally controlled processors, showing how they enhance connectivity, enable long-range entangling operations, and bridge distant regions without increasing local control overhead. We discuss physical implementations and architectural strategies illustrating how these elements extend the capabilities of global-control schemes. A complementary interpretation in terms of quantum batteries naturally emerges, connecting global-control architectures with concepts from quantum thermodynamics while highlighting the distinct operational role of quantum actuators.
\end{abstract}

\maketitle

\section{Introduction}

Global-control quantum computation~\cite{Lloyd_1993, benjamin_2000} offers a promising route toward scalable quantum information processing by reducing the need for fine-grained local control.  
Instead of individually addressing each qubit, these architectures rely on spatially uniform control fields combined with static structures that determine where and how interactions are activated~\cite{benjamin_2001_2, benjamin_2003, benjamin-bose_2004, cesa2023, menta2024globally, cioni2024conveyorbelt, menta2025building, hu2025universal, aiudi2026}.  
In this paradigm, computation is driven by global pulses, while internal elements regulate the flow of quantum information and the activation of multi-qubit operations.

This approach directly addresses several key challenges faced by current quantum computing platforms, including dense control wiring, individual qubit addressability, fast classical feedback, and complex routing schemes, all of which scale unfavorably with system size~\cite{mohseni2024build}.  
By shifting complexity from external control hardware to internal device structure, global-control schemes can significantly simplify the physical implementation of large-scale quantum processors.  
However, this shift introduces a fundamental requirement: the presence of internal components capable of selectively enabling or inhibiting interactions without relying on local control lines.

In this work, we introduce the concept of {\it quantum actuators} as such internal components.  
Quantum actuators are auxiliary quantum systems embedded within the processor that mediate gate operations by conditionally activating interactions between computational qubits.  
They do not encode logical information and remain passive during circuit compilation, preserving coherence without requiring individual control.  
During operation, however, they enable selective multi-qubit gates, directional transfer of quantum information, and programmable connectivity within globally driven architectures.

We present a systematic framework for incorporating quantum actuators into globally controlled quantum processors.  
We show how they can enhance connectivity, enable long-range entangling operations, and bridge spatially separated regions of a device without increasing control complexity.  
This perspective naturally extends existing global-control architectures by promoting these auxiliary elements from static structural features to active, functional components of the computation.

A key observation is that quantum actuators admit a natural energetic interpretation.  
Their operation can be viewed in terms of transient energy exchange with the global control fields: during a gate operation, an actuator effectively stores and releases interaction energy, enabling or suppressing couplings between qubits.  
In this sense, quantum actuators share conceptual similarities with {\it quantum batteries}~\cite{Alicki2013, RevModPhys.96.031001, Ferraro2026}, which are quantum systems designed to store and transfer energy in a controlled manner~\cite{PhysRevX.5.041011, ferraro2018high,
Campaioli2017, Andolina2019, caravelli2021energy, Andolina2025, canzio2025extracting, charging-prl2026}.  
As we will discuss below, however, quantum actuators differ in their operational role, as they are not used to implement arbitrary quantum transformations but rather to selectively activate or suppress specific interactions~\cite{Schirmer2008, Zhang2014, Aiello2015, Layden2016, cesa2026engineering, white2026quantum}.  
Here, this interpretation is not assumed as a starting point, but rather emerges as a complementary viewpoint that connects global-control architectures with ideas from quantum thermodynamics.

The remainder of this paper is organized as follows.  
In Sec.~\ref{sec:global} we review globally driven quantum computing architectures and the principles underlying their operation.  
In Sec.~\ref{sec:batteries} we introduce quantum actuators, discuss their physical implementation, and formalize their role in mediating gate operations.  
In Sec.~\ref{sec:applications} we present applications to programmable architectures and modular quantum processors based on ladder-like and conveyor-belt geometries.  
Finally, Sec.~\ref{sec:discussion} provides a summary and discusses the energetic interpretation of quantum actuators in relation to quantum batteries.

\section{Global quantum computation}\label{sec:global}

Globally controlled quantum computation~\cite{Lloyd_1993, benjamin_2000} has emerged as a promising strategy to mitigate control scalability limitations and the so-called ``wiring problem'' in superconducting quantum processors. Rather than relying on individually addressed qubits, these architectures employ uniform global control fields combined with engineered static inhomogeneities that determine where and how quantum operations occur. In this section, we briefly review the physical principles underlying global-control quantum computation and introduce the notation and definitions that will be used throughout the paper, focusing on the superconducting architectures proposed in Refs.~\cite{menta2024globally, cioni2024conveyorbelt, menta2025building}.

\subsection{Model and global-control structure}

\begin{figure}[t!]
    \centering
    \includegraphics[width=0.7\columnwidth]{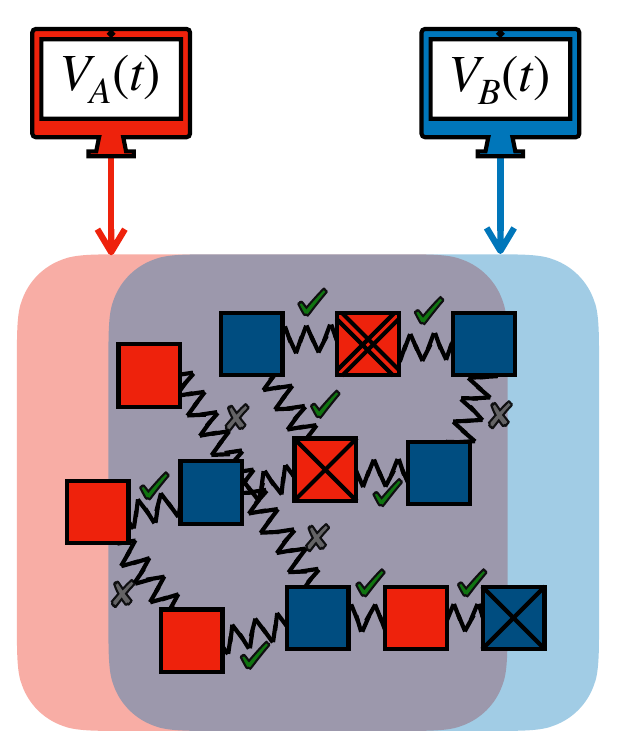}
    \caption{Example of a globally driven quantum architecture with two species ($\#\mathcal{S}=2$), $A$ and $B$. 
    Qubits are arranged on a random illustrative planar graph with always-on nearest-neighbor ZZ interactions, 
    while species assignment satisfies constraints (P1) and (P2): no two adjacent qubits belong 
    to the same species, and all qubits of a given species are driven by the same global control field. Different colors indicate distinct species, and edges represent static ZZ couplings.}
    \label{fig:example}
\end{figure}

In the globally driven ZZ-based architectures considered in Refs.~\cite{menta2024globally, cioni2024conveyorbelt, menta2025building}, a set of qubits is arranged on a planar interaction graph. Edges of the graph represent always-on nearest-neighbor ZZ couplings of uniform strength~$\zeta$. The qubits are partitioned into a finite set $\mathcal{S}$ of distinct species, labeled by an index $\chi$, such that no two neighboring qubits belong to the same species, as illustrated in Fig.~\ref{fig:example}. This construction satisfies two essential design principles~\cite{menta2025building}:
\begin{itemize}
\item[(P1)] each qubit interacts exclusively with nearest neighbors of different species;
\item[(P2)] all qubits belonging to a given species $\chi$ are driven collectively by the same classical control signal
\begin{equation}
V_{\chi}(t) := {\cal A}_{\chi}(t)\sin\!\big(\omega_{\mathrm{d},\chi} t + \phi_{\chi}(t)\big),
\end{equation}
generated by an independent electrical source characterized by a fixed carrier frequency $\omega_{\mathrm{d},\chi}$ and, in general, time-dependent amplitude ${\cal A}_{\chi}(t)$ and phase $\phi_{\chi}(t)$.
\end{itemize}
As a consequence, the number of independent control channels scales with the number of species rather than with the total number of qubits. Moreover, by construction, qubits belonging to the same species do not interact directly.

We now introduce the notation used throughout this work. Let $\hat{\sigma}_i^{(x,y,z)}$ denote the Pauli operators acting on the $i$-th qubit, defined in the local energy eigenbasis $\{\ket{g_i},\ket{e_i}\}$. The total Hamiltonian of the system takes the form $\hat{H}(t) := \hat{H}_0 + \hat{H}_{\rm drive}(t)$ where the static contribution describes both the local qubit energies and the always-on ZZ interactions,
\begin{equation}
\hat{H}_0 :=
\sum_{\chi \in \mathcal{S}} \sum_{i \in \chi}
\frac{\hbar \omega_i}{2}\hat{\sigma}_i^{(z)}
+
\sum_{\langle i,j \rangle}
\frac{\hbar \zeta}{2}
\hat{\sigma}_i^{(z)} \otimes \hat{\sigma}_j^{(z)} .
\label{H0}
\end{equation}
The driving term accounts for the action of the global control fields and reads
\begin{eqnarray}
\hat{H}_{\rm drive}(t)
&=&
\sum_{\chi \in \mathcal{S}} V_{\chi}(t)
\sum_{i \in \chi} {\cal V}_i \hat{\sigma}_i^{(y)}
\label{Hdrive}
\end{eqnarray}
where ${\cal V}_i$ denotes the coupling strength between the $i$-th qubit and its associated drive field. The corresponding Rabi frequency is given by
\begin{equation}
\Omega_{\chi,i}(t) := \frac{{\cal A}_{\chi}(t)\,{\cal V}_i}{\hbar}.
\label{rabi}
\end{equation}

For each species $\chi$, the qubit transition frequencies $\omega_i$ are detuned from the drive frequency $\omega_{\mathrm{d},\chi}$ by an amount proportional to the interaction strength~$\zeta$, $\omega_i = \omega_{\mathrm{d},\chi} + \kappa_i \zeta$, where $\kappa_i$ denotes the coordination number of site $i$, i.e., the number of nearest neighbors coupled to qubit $i$.

A key architectural ingredient is that the coupling constants ${\cal V}_i$ can be engineered at the fabrication stage~\cite{menta2024globally}. This allows each species $\chi$ to be partitioned into two disjoint subsets: a set $\chi^{\rm r}$ of regular qubits with a fixed reference coupling strength $\bar{\cal V}$, and a set $\chi^{\times}$ of crossed qubits with twice that coupling, the so-called “crossed-qubit'' method~\cite{menta2025building}. As a result, the corresponding Rabi frequencies satisfy $\Omega_{\chi^{\rm r}}(t) := {\cal A}_{\chi}(t)\bar{\cal V}/\hbar$ and $\Omega_{\chi^{\times}}(t):= 2\,\Omega_{\chi^{\rm r}}(t)$.
Apart from these fabrication-level distinctions, all control parameters remain uniform within each species. The crossed-qubit mechanism effectively introduces localized control features while preserving the global nature of the driving fields.

\subsection{Effective control dynamics}

The computational power of the globally controlled architectures described above relies on the concatenation of ordered sequences of global unitary transformations of the form
\begin{equation}
\hat{U}_{\rm seq} := \hat{U}_{\chi_n} \cdots \hat{U}_{\chi_2} \hat{U}_{\chi_1},
\label{defseq}
\end{equation}
where each unitary $\hat{U}_{\chi_j}$ corresponds to the activation of a specific species over a designated time interval. When combined with suitably encoded initial states, these sequences enable the transport and manipulation of logical quantum information through the lattice, exploiting engineered inhomogeneities such as crossed elements~\cite{menta2024globally, cioni2024conveyorbelt}.

To elucidate the structure of the effective dynamics, it is convenient to move to a rotating frame defined by the unitary transformation $\hat{U}_{\rm rf}(t) := \bigotimes_i
e^{i \hat{\sigma}_i^{(z)} \omega_{\mathrm{d},i} t / 2 }$, 
and to apply the rotating-wave approximation (RWA), neglecting rapidly oscillating terms at frequencies $\omega_i + \omega_{\mathrm{d},\chi}$. In this frame, the Hamiltonian reduces to~\cite{menta2024globally}
\begin{eqnarray}
\hat{H}_{\rm rf}(t)
&\simeq&
\sum_{\chi \in \mathcal{S}} \sum_{i \in \chi}
\frac{\hbar \Omega_{\chi,i}(t)}{2}
\Big[
e^{i\phi_{\chi}(t)} \ket{g_i}\bra{e_i} + {\rm h.c.}
\Big]
\nonumber \\
&&+
\sum_{\langle i,j \rangle}
2\hbar\zeta\, \ket{e_i e_j}\bra{e_i e_j}.
\label{Hfinal}
\end{eqnarray}

A crucial operational condition is the strong-coupling, or dynamical blockade~\cite{riccardi2026}, regime $\eta_{\rm BR} := \left| \frac{\zeta}{\Omega_{\chi^{\rm r}}} \right| \gg 1 $
which gives rise to a conditional suppression of local nearest neighboring rotations. In this regime, if a qubit has at least one neighboring qubit in the excited state, the applied drive is effectively unable to induce transitions on that qubit.
The resulting effective evolution during a control interval ${\cal T}$ can be expressed in terms of conditional unitary operators
\begin{equation}
\hat{W}_j(\theta_j,\bm{n}_j)
:= \hat{\openone}_j \otimes \hat{Q}_{\langle j \rangle}
+ \hat{\mathbb{R}}_j(\theta_j,\bm{n}_j) \otimes \hat{P}_{\langle j \rangle},
\label{def_W}
\end{equation}
where $\hat{P}_{\langle j \rangle}$ projects onto the subspace in which all neighbors of qubit $j$ are in the ground state, and $\hat{Q}_{\langle j \rangle} = \hat{\openone}_{\langle j \rangle} - \hat{P}_{\langle j \rangle}$ projects onto the complementary blocked subspace. The operator
\begin{eqnarray}
\hat{\mathbb{R}}_j(\theta_j,\bm{n}_j)
&:=&
\exp\!\left[-\frac{i\theta_j}{2}\,\bm{n}_j \cdot \vec{\sigma}_j\right]
 \\
&=&
\overset{\longleftarrow}{\exp}\!\left(
-\frac{i}{2}\int_{\cal T} dt\, \Omega_j(t)
\big[e^{i\phi_j(t)}\ket{g_j}\bra{e_j} + {\rm h.c.}\big]
\right) \nonumber
\label{identityR}
\end{eqnarray}
denotes a single-qubit rotation by angle $\theta_j$ around the unit vector $\bm{n}_j$.
Because no two neighboring qubits belong to the same species and the control fields act collectively on each species, activating a single control $V_\chi(t)$ induces identical transformations on all qubits of type $\chi$. Taking into account the crossed-qubit structure, the resulting global unitary operation can be written as
\begin{eqnarray}
\hat{U}_\chi
&:=&
\left.
\overset{\longleftarrow}{\exp}
\left(
-\frac{i}{\hbar}\int_{\cal T} dt\, \hat{H}_{\rm rf}(t)
\right)
\right|_{\eta_{\rm BR}\gg 1}
\nonumber \\
&=&
\hat{W}_{\chi^{\rm r}}(\theta^{\rm r},\bm{n}^{\rm r})
\hat{W}_{\chi^{\times}}(\theta^{\times},\bm{n}^{\times}),
\label{pulsechi}
\end{eqnarray}
with
\begin{eqnarray}
\hat{W}_{\chi^{\rm r}}(\theta^{\rm r},\bm{n}^{\rm r})
&:=&
\prod_{i \in \chi^{\rm r}}
\hat{W}_i(\theta^{\rm r},\bm{n}^{\rm r}),
\nonumber \\
\hat{W}_{\chi^{\times}}(\theta^{\times},\bm{n}^{\times})
&:=&
\prod_{i \in \chi^{\times}}
\hat{W}_i(\theta^{\times},\bm{n}^{\times}).
\label{pulsechidef}
\end{eqnarray}

Despite possible correlations between the rotation parameters of the regular and crossed subsets, it has been shown that by suitably shaping the global pulses $V_\chi(t)$ one can realize independent control-unitary operations on $\chi^{\rm r}$ and $\chi^{\times}$~\cite{menta2025building}. This effectively increases the number of available control degrees of freedom without introducing additional control lines.

Finally, since the subsets $\chi^{\rm r}$ and $\chi^{\times}$ are disjoint and no ZZ interactions connect qubits of the same species, the corresponding transformations commute,
\begin{equation}
\big[
\hat{W}_{\chi^{\rm r}}(\theta^{\rm r},\bm{n}^{\rm r}),
\hat{W}_{\chi^{\times}}(\theta^{\times},\bm{n}^{\times})
\big] = 0 .
\label{commute}
\end{equation}
For fixed $\chi$, the set of operators $\hat{W}_{\xi}(\theta,\bm{n})$ with $\xi \in \{\chi^{\rm r},\chi^{\times}\}$ forms a representation of $SU(2)$, with composition and inversion rules identical to those of single-qubit rotations. This algebraic structure underlies the computational capabilities of the global-control architectures reviewed here. It is worth noting that, as shown in Ref.~\cite{menta2025building}, the crossed-qubit method can be generalized to engineer, for instance, double-crossed qubits with a distinct Rabi frequency, defined as $\Omega_{\chi^{\mathbb{X}}}(t) := 2\,\Omega_{\chi^{\times}}(t)$. This generalization allows for reduced qubit overhead by enabling more effective local addressing.

\subsection{Globally driven geometries}

It has been shown~\cite{menta2024globally, cioni2024conveyorbelt, menta2025building} that, under the global-control principles (P1) and (P2), two distinct classes of geometries support universal quantum computation. We refer to these as {\it ladder-like architectures} and {\it conveyor-belt-like architectures}. The former corresponds to a quasi-two-dimensional layout, while the latter is a one-dimensional geometry in which qubits are arranged along a closed, deformed loop. In the following, we briefly review the main operational principles underlying these two constructions.

\subsubsection{Ladder-like architectures}

A ladder-like architecture encoding $N$ computational qubits consists of $N$ rows of physical qubits. Each row contains $2N+3$ physical qubits arranged in a periodic pattern of species $\mathcal{S}$ that satisfies the global-control constraints. Adjacent rows are connected by additional {\it coupler} qubits—implemented as crossed qubits—that function as the ``legs'' of the ladder. There are $N-1$ such inter-row crossed qubits, which define dedicated computational regions where two-qubit gates can be implemented. In addition, each row contains one crossed qubit that enables the execution of single-qubit gates. A schematic representation of this geometry is shown in Fig.~\ref{fig:ladder-cb}(a). The total number of physical qubits therefore scales as $\mathcal{O}(N^2)$, as summarized in Table~\ref{tab1}.

This overhead in physical qubits is required to encode and manipulate logical quantum information, which is stored in a specific column of the ladder geometry known as the {\it Information Carrying Column} (ICC). The ICC is defined as the interface between two classical product states: the N\'eel-ordered state $\ket{\rm N} := \vert \cdots g e g e g \rangle$ and the ferromagnetic state $\ket{\rm F} := \vert g g g g \cdots \rangle$. If the ICC is located on the $k$-th row, the corresponding encoded state takes the form
\begin{equation}
\ket{{\rm ICC}_k} := \ket{\rm N}^{\otimes N} \otimes \ket{\Psi}_k \otimes \ket{\rm F}^{\otimes N} \ ,
\end{equation}
where $\ket{\Psi}_k = \ket{\psi_1} \otimes \cdots \otimes \ket{\psi_N}$ encodes the logical quantum information.
By concatenating global unitaries $\hat{U}_{\rm shift}$ of the form given in Eq.~\eqref{defseq}, the ICC can be translated along the ladder, enabling controlled transport of logical information. In particular, one has $\hat{U}_{\rm shift}^{\ell} \ket{{\rm ICC}_k} = \ket{{\rm ICC}_{k+\ell}}$ and $\hat{U}_{\rm shift}^{\dagger \ell} \ket{{\rm ICC}_k} = \ket{{\rm ICC}_{k-\ell}}$. This ability to shift the ICC, combined with the possibility of selectively addressing specific regions of the architecture where crossed (or double-crossed) qubits are located~\cite{menta2025building}, allows for the implementation of single-qubit and two-qubit gates through global control operations of the form~\eqref{pulsechi}. Together, these ingredients enable universal quantum computation within ladder-like architectures~\cite{menta2024globally, menta2025building}.

\subsubsection{Conveyor-belt-like architectures}

The conveyor-belt-like architecture, originally proposed in Ref.~\cite{cioni2024conveyorbelt},
provides a globally driven platform for universal quantum computation with a linear overhead,
requiring only $\mathcal{O}(N)$ physical qubits to encode $N$ computational qubits, as summarized
in Table~\ref{tab1}. 
In this geometry, qubits are arranged along a closed one-dimensional loop, and logical
information is encoded in delocalized \textit{information-carrying qubits} (ICQs),
denoted by $Q_1,\ldots,Q_N$.

Between each pair of neighboring computational qubits $Q_i$ and $Q_{i+1}$,
additional physical qubits are inserted and act as mediators of the global dynamics.
We denote these ancillary blocks by $S_{i,i+1}$, with $S_{N,1}$ closing the loop.
In the original proposal of Ref.~\cite{cioni2024conveyorbelt}, each block $S_{i,i+1}$
consists of three physical qubits encoding alternating classical patterns
$\ket{\rm F/N}:=\ket{ggg/geg}$, while in the low-overhead variant of
Ref.~\cite{menta2025building} each block is reduced to a single physical qubit.

With this notation, the logical state of the processor can be written as
\begin{eqnarray}\nonumber &\ket{\rm ICS}& := \sum_{\vec{x}\in \{ g,e\}^N} \alpha_{\vec{x}} |x_1\rangle_{Q_1} |{\rm R} \rangle_{S_{1,2}} |x_2\rangle_{Q_2} |{\rm R}\rangle_{S_{2,3}} |x_3\rangle_{Q_3} |{\rm R}\rangle_{S_{3,4}} \nonumber \\ &\cdots& |x_{N-1}\rangle_{Q_{N-1}} |{\rm R} \rangle_{S_{N-1,N}} |x_N\rangle_{Q_N} |{\rm R} \rangle_{S_{N,1}} \otimes |g\rangle_{A^{\times}}, \label{eq:encoding} \end{eqnarray}
where the register states $\ket{\rm R}_{S_{i,i+1}}$ alternate between
$\ket{\rm F}$ and $\ket{\rm N}$ in the scheme of Ref.~\cite{cioni2024conveyorbelt},
or reduce to a fixed ground state in the low-overhead realization
of Ref.~\cite{menta2025building}. 
The final qubit $\ket{g}_{A^{\times}}$ does not belong to the loop; it is coupled
to three computational qubits and enables the implementation of a one-shot
Toffoli gate, thereby breaking parity constraints.

The core operational principle of the conveyor-belt architecture is the ability
to perform global SWAP operations between neighboring computational qubits using
alternating sequences of species-selective global unitaries of the form
Eq.~(\ref{defseq}). 
By concatenating these global SWAP operations, logical information can be
transported coherently along the loop, either clockwise or counterclockwise.
For instance, one finds
\begin{equation}
\hat{U}^{\circlearrowright_e,\circlearrowleft_o}_{\rm swap}
\ket{x_1 x_2 x_3 \ldots x_N}
=
\ket{x_N x_3 x_2 \ldots x_1},
\end{equation}
where $\circlearrowright_e$ ($\circlearrowleft_o$) denotes a clockwise
(anticlockwise) SWAP applied to even (odd) indexed qubits.

The combination of directional transport via global SWAPs, the presence of a
crossed qubit within the loop, and the auxiliary coupler qubit allows
to implement arbitrary single- and three-qubit computational gates, thereby enabling
universal quantum computation under global control~\cite{cioni2024conveyorbelt,menta2025building}.

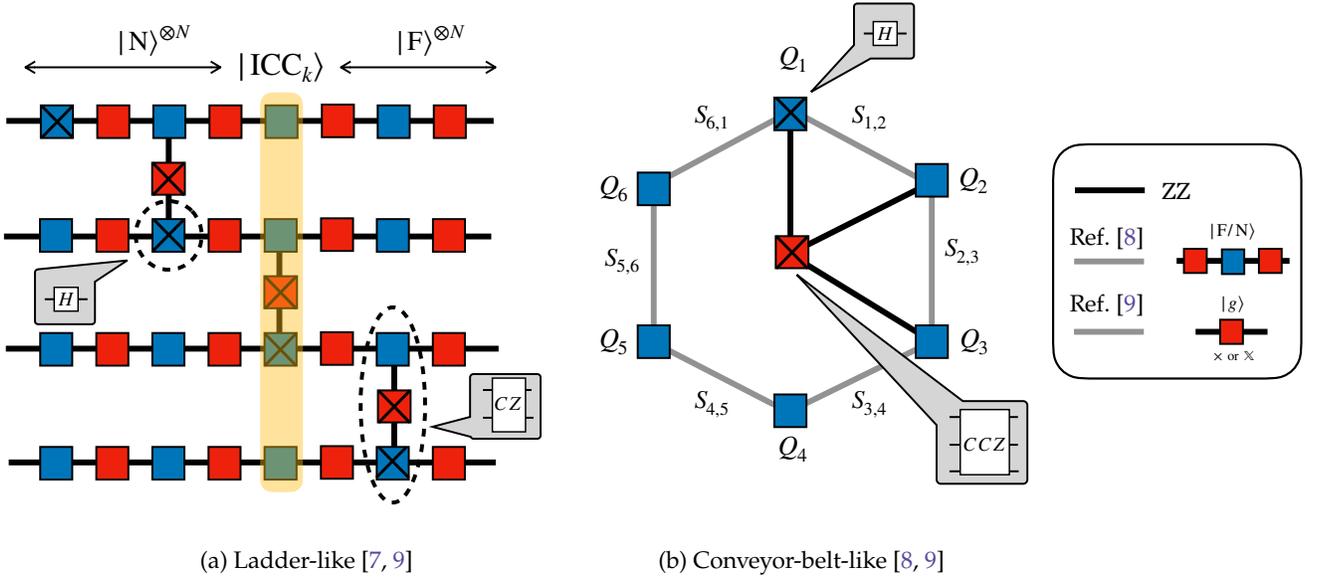
\begin{figure*}
    \centering
    \begin{overpic}[width=1.0\linewidth]{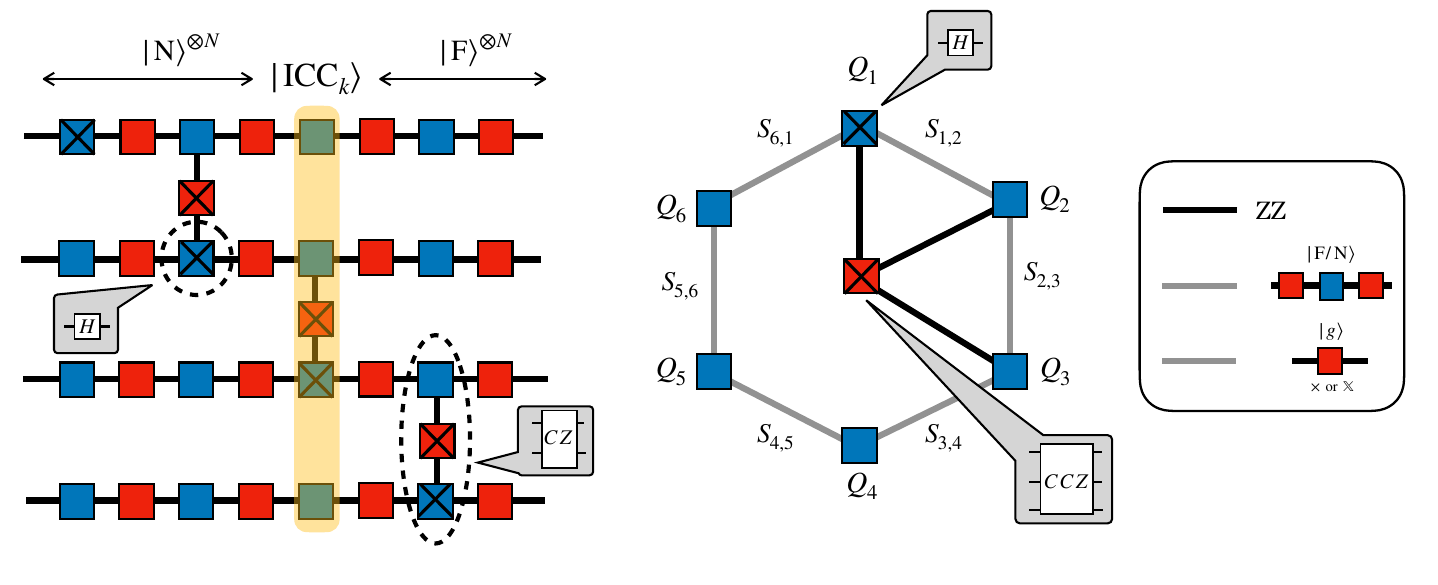}
        \put(16, -2){(a) Ladder-like~\cite{menta2024globally, menta2025building}}
        \put(50, -2){(b) Conveyor-belt-like~\cite{cioni2024conveyorbelt, menta2025building}}
        \put(80.5, 22){Ref.~\cite{cioni2024conveyorbelt}}
        \put(80.5, 17){Ref.~\cite{menta2025building}}
        \end{overpic}
    \vspace{0.5cm}
    \caption{Schematic globally driven architectures supporting universal quantum computation under constraints (P1) and (P2).
    (a) Ladder-like architecture: logical information is encoded in an ICC separating ferromagnetic ($\ket{\rm F}$) and N\'eel ($\ket{\rm N}$) background phases. For simplicity, a two-species implementation is shown and double-crossed qubits are omitted. Global control sequences shift the ICC along the ladder, while crossed qubits (marked by $\times$ or $\mathbb{X}$) enable effective local single- and two-qubit gates via the blockade mechanism. (b) Conveyor-belt-like architecture: computational qubits $Q_i$ form a closed loop and are separated by auxiliary segments $S_{i,i+1}$ storing classical reference states. Alternating global pulses implement nearest-neighbor SWAP operations, allowing logical qubits to circulate around the loop. A crossed element and a parity-breaking CZZ gate complete the set of resources required for universality. Static ZZ couplings are indicated by solid links throughout.}
    \label{fig:ladder-cb}
\end{figure*}

Figure~\ref{fig:ladder-cb} illustrates both examples of superconducting globally driven just presented. The exploration of novel topologies and circuit geometries remains an active area of research.

\begin{table}[!htp]
\centering
\begin{tabular}{c|cccc}
& drive lines & phys. qbits & $\times$-qbits & $\mathbb{X}$-qbits \\
\hline
ladder~\cite{menta2024globally} & 3  & $2N^2 + 4N - 1$  & $3N-1$ & $\varnothing$  \\
conveyor-belt~\cite{cioni2024conveyorbelt} & 2 & $4N + 1$ & 2 & $\varnothing$ \\
ladder~\cite{menta2025building} & 2 & $2N^2 + 4N - 1$ & $\frac{N}{2} -1$ & $N+ \frac{N}{2}$ \\
conveyor-belt~\cite{menta2025building} & 2 & $2N + 1$ & $N$ & 2 \\
\end{tabular}
\caption{Summary of the different configurations of species, physical qubits, and crossing elements for the architectures of Refs.~\cite{menta2024globally, cioni2024conveyorbelt, menta2025building, aiudi2026} $N$ is the number of computational qubits, assumed here to be even and $N\geq2$.}
\label{tab1}
\end{table}

\subsection{Multi-qubit gates}

It is worth elaborating on how entangling two- and three-qubit gates, such as the controlled-$Z$ ($CZ$) and controlled-controlled-$Z$ ($CCZ$) gates, are implemented within the globally driven architectures reviewed above.
These operations rely on a dynamical blockade mechanism~\cite{riccardi2026}, combined with the crossed-qubit method extensively discussed in Ref.~\cite{menta2025building}.
Importantly, they arise as {\it native} operations of the architecture and do not require decompositions into elementary two-qubit gates.

Consider a crossed qubit $A^{\times}$ coupled via always-on ZZ interactions to a set of neighboring qubits $\{B_k\}$.
In the rotating frame and under the blockade condition $\eta_{\rm BR} \gg 1$, the effective Hamiltonian takes the form
\begin{eqnarray}
\hat{H}_{\rm eff}(t)
&=&
\frac{\hbar \Omega_{A^{\times}}(t)}{2}
\Big[
e^{i\phi(t)} |g\rangle_{A^{\times}}\langle e| + {\rm h.c.}
\Big]
\otimes \hat{P}_{\langle A^{\times}\rangle}
\nonumber \\
&&+
\hat{H}_{\rm blocked}\otimes \hat{Q}_{\langle A^{\times}\rangle},
\end{eqnarray}
where $\hat{P}_{\langle A^{\times}\rangle}$ projects onto the subspace in which all neighboring qubits $\{B_k\}$ are in the ground state $|g\rangle$, while $\hat{Q}_{\langle A^{\times}\rangle}=\openone-\hat{P}_{\langle A^{\times}\rangle}$ projects onto the complementary (blocked) subspace.
In the blocked subspace, transitions of $A^{\times}$ are suppressed by the large interaction-induced detuning.

We now apply a $2\pi$ pulse to the crossed qubit,
\begin{equation}
\hat{\mathcal{Z}}_{A^{\times}}
:=
\hat{W}_{A^{\times}}(2\pi,\bm{n})
\end{equation}
Let us analyze its action on a generic basis state
\begin{equation}
|\Psi\rangle
=
|a\rangle_{A^{\times}}
\otimes
|b_1 b_2 \ldots b_m\rangle_{B_1 B_2 \ldots B_m},
\quad b_k\in\{g,e\}.
\end{equation}

If all neighboring qubits are in the ground state, $b_k=g$ for all $k$, then $\hat{P}_{\langle A^{\times}\rangle}|\Psi\rangle=|\Psi\rangle$.
In this case, the crossed qubit undergoes a full $2\pi$ rotation,
\begin{equation}
\hat{W}_{A^{\times}}(2\pi,\bm{n})|g\rangle_{A^{\times}}
=
-\,|g\rangle_{A^{\times}},
\end{equation}
independently of the rotation axis $\bm{n}$.
Therefore, $|\Psi\rangle \to
-\,|\Psi\rangle$. If at least one neighboring qubit is excited, there exists a $k$ such that $b_k=e$, and hence
$\hat{Q}_{\langle A^{\times}\rangle}|\Psi\rangle=|\Psi\rangle$.
In this case, the effective Rabi frequency vanishes and the crossed qubit does not evolve.
The state remains unchanged: $|\Psi\rangle \to|\Psi\rangle$.
Combining the two cases, the net action of $\hat{\mathcal{Z}}_{A^{\times}}$ is
\begin{equation}
\hat{\mathcal{Z}}_{A^{\times}}
=
-\,\hat{P}_{\langle A^{\times}\rangle}
+
\hat{Q}_{\langle A^{\times}\rangle}.
\end{equation}
Up to a global phase, this operator applies a $(-1)$ phase factor if and only if all neighboring qubits are in the ground state.

\paragraph{CZ and CCZ gates.}
For two neighboring qubits $B_1$ and $B_2$, this realizes a $CZ$ gate in the computational basis
\begin{equation}
\hat{\mathcal{Z}}_{A^{\times}}
\;\equiv\;
{\rm diag}(1,1,1,-1)_{B_1 B_2},
\end{equation}
while for three neighboring qubits $B_1,B_2,B_3$, it implements a $CCZ$ gate,
\begin{equation}
\hat{\mathcal{Z}}_{A^{\times}}
\;\equiv\;
{\rm diag}(1,\ldots,1,-1)_{B_1 B_2 B_3}.
\end{equation}
These multi-qubit controlled-phase gates arise from a {\it single global pulse} applied to the crossed qubit and therefore constitute native, hardware-efficient entangling operations within globally driven architectures.

\section{Global control via quantum actuators}\label{sec:batteries}

In this section, we introduce the concept of {\it quantum actuators}~\cite{Zhang2014}. 
Building on the global-control architectures reviewed above, we propose new variants in which these devices are integrated as functional elements of the quantum processor.

\subsection{Quantum actuators}

We begin by introducing the notion of a quantum actuator from an operational and architectural perspective.

\begin{tcolorbox}
\begin{definition}[Quantum actuator]
A quantum actuator (QA) is an auxiliary quantum system that mediates the implementation of some quantum gates between computational qubits in a quantum processor.
The actuator does not encode logical quantum information and does not participate directly in the computational register.
Instead, its internal state conditions or enables specific physical interactions—such as blockade effects or conditional phase accumulation—between neighboring qubits, thereby controlling the execution of some quantum gates.
\end{definition}
\end{tcolorbox}

Quantum actuators play a role analogous to classical control elements in conventional hardware architectures: they regulate when and how interactions occur, while remaining external to the logical computation itself.

Within the present framework, the physical realization of the actuator can be extremely simple.
In the minimal case, it may consist of a single physical qubit described by the Hamiltonian
\begin{equation}
\hat{H}_{\rm QA} = \frac{\hbar \omega_{\rm QA}}{2}\,\hat{\sigma}_{\rm QA}^{(z)} .
\end{equation}
Despite its simplicity, such a system can transiently store energy injected by global control fields and release it in a controlled manner to mediate multi-qubit gate operations.
In practice, charging the actuator corresponds to a unitary operation, such as a $\pi$ pulse implementing the transition $\ket{g}_{\rm QA} \to \ket{e}_{\rm QA}$.

\subsection{Integration within quantum processors}

We now introduce two conceptual steps that allow quantum actuators to be embedded within globally driven quantum processors.
These steps provide the logical bridge between the architectures reviewed in the previous sections and the variants proposed here.
Each step is supported by a concrete dynamical realization based on blockade mechanisms.

\begin{tcolorbox}
\begin{conceptual}[Crossed qubit as a quantum actuator]\label{conceptual1}
The first conceptual step consists in reinterpreting the crossed qubit $A^{\times}$ as a quantum actuator.
Rather than viewing $A^{\times}$ as a computational element, we regard it as an auxiliary quantum system whose sole function is to mediate multi-qubit gate operations.
From an energetic perspective, this actuator transiently stores energy supplied by global control fields and releases it to condition the activation of interaction terms responsible for two- and three-qubit gates.
\end{conceptual}
\end{tcolorbox}

\begin{figure*}[t]
    \centering
    \begin{overpic}[width=1.0\linewidth]{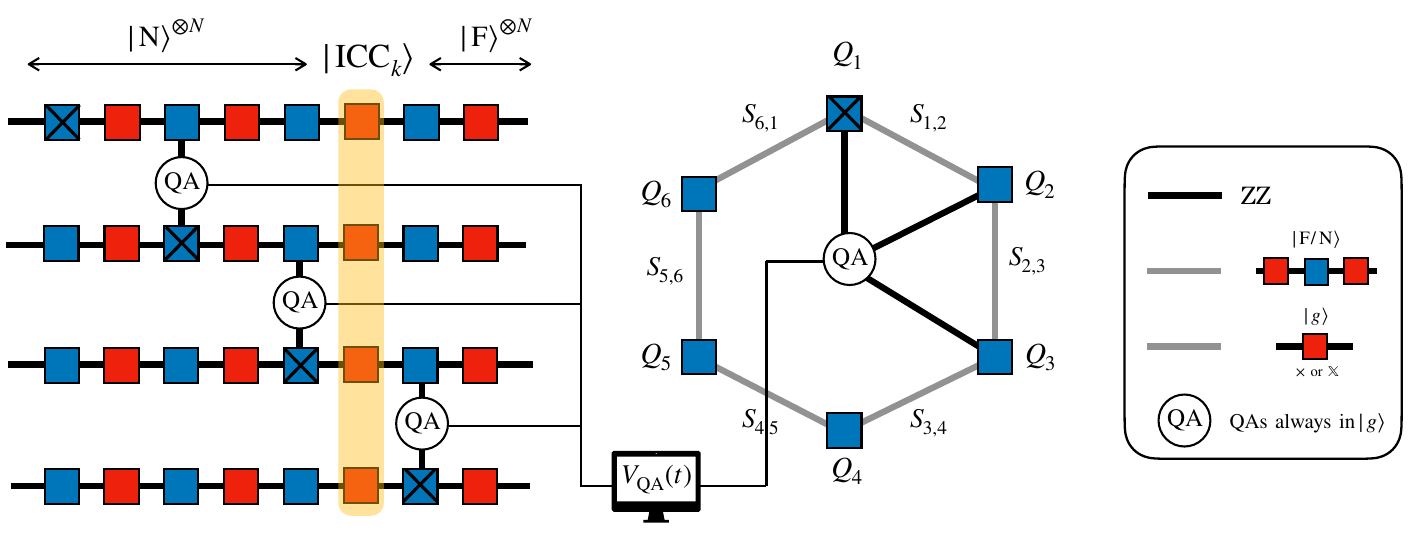}
        \put(14, -4){(a) Ladder with QAs}
        \put(52, -4){(b) Conveyor-belt with QA}
        \put(80.5, 21){Ref.~\cite{cioni2024conveyorbelt}}
        \put(80.5, 15.5){Ref.~\cite{menta2025building}}
    \end{overpic}
    \vspace{0.5cm}
    \caption{Variants of globally driven architectures inspired by Refs.~\cite{menta2024globally, cioni2024conveyorbelt, menta2025building}, augmented with quantum actuators operating under global control—see conceptual step~\autoref{conceptual1}.
    (a) Ladder-like architecture, with $N-1$ quantum actuators for an $N$-qubit ICC.
    (b) Conveyor-belt-like architecture, in which a single quantum battery enables effective three-qubit operations for any even $N \geq 2$.}
    \label{fig:variants}
\end{figure*}
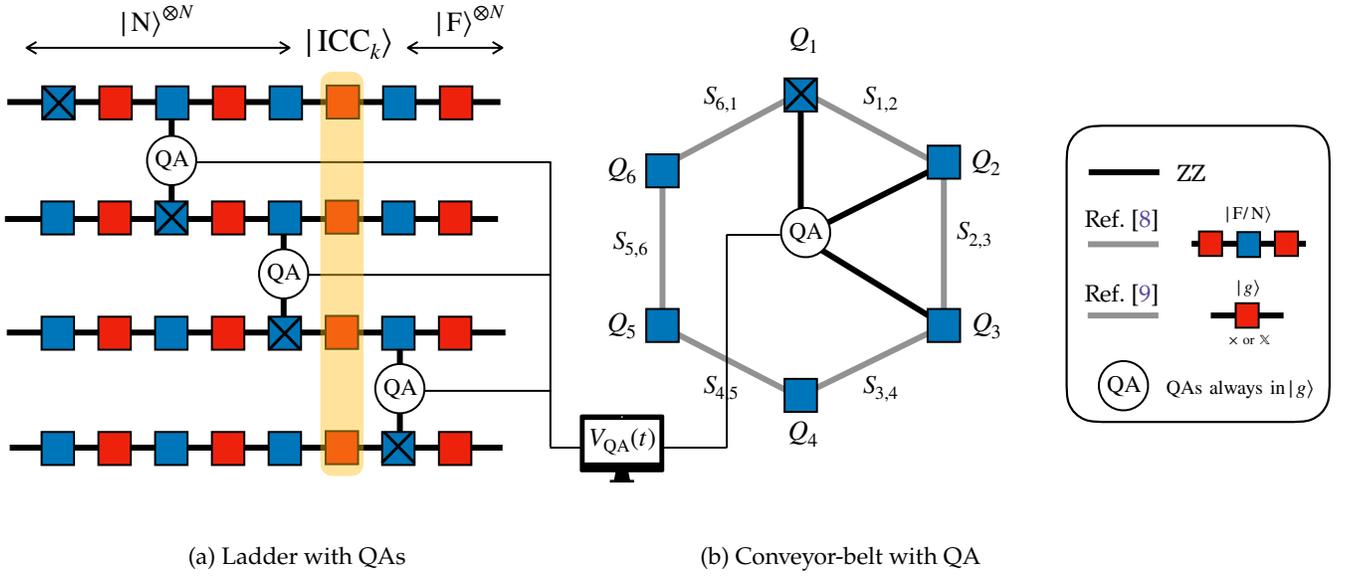

\paragraph{Formal realization.}
Consider a quantum actuator (QA) coupled via static ZZ interactions to a set of neighboring qubits $\{B_j\}_{j=1}^m$.
The system Hamiltonian reads $\hat{H}(t) = \hat{H}_0 + \hat{H}_{\rm drive}(t)$, with
\begin{equation}
\hat{H}_0 = \hat{H}_{\rm QA}
+ \sum_{j=1}^{m}\frac{\hbar\omega_{B_j}}{2}\hat{\sigma}^{(z)}_{B_j}
+ \sum_{j=1}^{m}\frac{\hbar\zeta}{2}\hat{\sigma}^{(z)}_{\rm QA}\hat{\sigma}^{(z)}_{B_j},
\end{equation}
and a drive acting on the QA,
\begin{equation}
\label{eq:drive-battery}
\hat{H}_{\rm drive}(t)
=
\hbar\Omega_{\rm QA}(t)\sin(\omega_{\rm d}t+\phi)\,\hat{\sigma}^{(y)}_{\rm QA}.
\end{equation}

Preparing the QA in its ground state and moving to the rotating frame, a $2\pi$ pulse under the rotating-wave approximation yields the unitary
\begin{equation}
\hat{U}
=
\hat{\openone}_{\rm QA} \otimes
\big(
\hat{\openone}_{B}
-
2\,\hat{P}_{\langle A \rangle}
\big),
\end{equation}
where $\hat{P}_{\langle A \rangle}$ projects onto the subspace in which all qubits coupled to the QA are in the ground state.
This corresponds to a conditional phase gate acting on the neighboring qubits, while the QA itself does not encode logical information.
In this sense, the crossed qubit functions as a quantum battery whose controlled charge--discharge enables a native multi-qubit gate.

\begin{tcolorbox}
\begin{conceptual}[Dynamical blockade via distributed quantum actuators]\label{conceptual2}
The second conceptual step consists in associating each physical qubit of the processor with an auxiliary quantum actuator.
These actuators are globally driven and maintained in an excited state, thereby acting as a distributed energetic resource.
Blockade-type interactions between an actuator and its associated qubit conditionally inhibit or release the stored energy, locally freezing or activating computational dynamics without requiring local control fields.
\end{conceptual}
\end{tcolorbox}

\paragraph{Freezing mechanism.}
Let $B$ denote a physical qubit coupled to an auxiliary actuator qubit QA via a ZZ interaction,
\begin{equation}
\hat{H}_0 =
\frac{\hbar\omega_B}{2}\hat{\sigma}^{(z)}_B
+ \hat{H}_{\rm QA}
+ \frac{\hbar\zeta}{2}\hat{\sigma}^{(z)}_B\hat{\sigma}^{(z)}_{\rm QA}.
\end{equation}
A drive acting on qubit $B$ is given by Eq.~\eqref{eq:drive-battery}.

If the battery is initialized in the excited state $\ket{e}_{\rm QA}$, the transition frequency of $B$ is shifted by $\zeta$.
In the blockade regime $\lvert \zeta / \Omega_{B} \rvert \gg 1$, the drive becomes off-resonant and the effective evolution reduces to
\begin{equation}
\hat{U}_B \simeq \hat{\openone}_B ,
\end{equation}
so that the physical qubit is dynamically frozen.
The energetic state of the battery thus locally activates or deactivates computation while preserving global control.

\medskip

Having introduced quantum actuators as mediating elements for gate operations, we now make a conceptual observation: they admit a natural interpretation as {\it quantum batteries}~\cite{Alicki2013, RevModPhys.96.031001, Ferraro2026, PhysRevX.5.041011, ferraro2018high, Campaioli2017, Andolina2019, caravelli2021energy, Andolina2025, canzio2025extracting, charging-prl2026} within the framework of quantum thermodynamics.

From this perspective, the state of the actuator controls the activation or suppression of computational processes. 
Excitation or de-excitation supplies or withdraws interaction energy, thereby enabling conditional quantum gates without additional local control resources.

This interpretation differs from previous proposals in which quantum batteries are used to directly implement arbitrary quantum operations~\cite{PhysRevX.11.021014, Chiribella2022, Yang_2022, castellano2025exact}. 
In those settings, the energetic cost can be substantial and typically requires highly coherent battery states. 
In contrast, here actuators are not responsible for generating the full set of logical transformations, but are instead used selectively to activate or suppress specific interactions or modify the effective connectivity of the processor. 
As a result, their operation does not rely on highly nonclassical states and can be implemented using simple two-level systems driven by global control fields.

\subsection{Globally driven architectures with quantum actuators}

We now introduce globally driven architectures in which quantum actuators are integrated, extending the global-control processors reviewed in Sec.~\ref{sec:global}.  
The resulting variants are shown in Fig.~\ref{fig:variants} and are directly derived from the ladder-like and conveyor-belt-like architectures of Refs.~\cite{menta2024globally, cioni2024conveyorbelt, menta2025building}.  
Their resource requirements are summarized in Table~\ref{tab2}.

The central architectural modification is the introduction of a dedicated global control line addressing the quantum actuators.  
This control line does not act directly on the computational qubits, but instead drives the actuators between their ground and excited states.  
Through blockade-type interactions, the internal state of an actuator conditionally activates multi-qubit interaction terms, thereby enabling two- and three-qubit gate operations under global driving.

This design principle is rooted in the crossed-qubit method.  
As shown in Refs.~\cite{menta2024globally, menta2025building}, a crossed element $A^{\times}$ behaves as an effective additional control line when driven globally, despite being a static component of the processor.  
Interpreting $A^{\times}$ as a quantum actuator makes this correspondence explicit: the actuator stores energy supplied by the global drive and releases it conditionally, resulting in a controlled phase operation on the neighboring qubits.  
In this sense, quantum actuators promote static connectivity into programmable interaction control without introducing local addressability.

In the ladder-like variant shown in Fig.~\ref{fig:variants}(a), $N-1$ quantum actuators are placed at the inter-row junctions associated with the information-carrying column (ICC).  
Each actuator mediates conditional two-qubit operations between adjacent rows.  
As summarized in Table~\ref{tab2}, this construction requires a single additional global drive line compared to the original ladder architecture, while preserving the scaling of physical qubits.

In the conveyor-belt-like variant shown in Fig.~\ref{fig:variants}(b), a single quantum actuator replaces the crossed element responsible for parity breaking and multi-qubit operations in the original proposals of Refs.~\cite{cioni2024conveyorbelt, menta2025building}.  
The actuator mediates an effective $CCZ$ gate via a conditional $2\pi$ pulse, enabling universal quantum computation with only one additional global control line, as detailed in Table~\ref{tab2}.

\begin{table}[!htp]
\centering
\begin{tabular}{c|ccc}
& drive lines & phys. qbits & QAs \\
\hline
Fig.~\ref{fig:variants}(a):~\cite{menta2024globally} (\cite{menta2025building}) & 3 (2) + 1 & $2N^2 + 4N - 1$  & $N-1$   \\
Fig.~\ref{fig:variants}(b):~\cite{cioni2024conveyorbelt}  (\cite{menta2025building})& 2 + 1 & $4N \ (2N) + 1$ & 1 \\
\end{tabular}
\caption{Resource overhead for the globally driven architectures shown in Fig.~\ref{fig:variants}.  
For each variant, we list the number of global drive lines, physical qubits, and quantum actuators (QAs) required to encode $N$ computational qubits.  
We assume $N$ even, with $N \geq 2$.}
\label{tab2}
\end{table}

Overall, the architectures of Fig.~\ref{fig:variants} demonstrate that quantum actuators can be seamlessly integrated into existing globally driven processors.  
They provide a mechanism for selectively activating interactions and enabling native multi-qubit operations, while maintaining the defining advantages of global control: uniform driving, reduced wiring complexity, and scalable resource overhead.

In this sense, quantum actuators function as coherent {\it switches} for the computational dynamics: when in the ground state, they inhibit the activation of multi-qubit interactions, while when excited they enable conditional gate operations via the blockade mechanism.  
They do not store or process logical quantum information and are returned to their initial state after each operation.

All the architectural variants introduced here explicitly satisfy conceptual step~\ref{conceptual1}.  
Quantum actuators mediate gate operations by transiently storing energy supplied by global drives, while remaining passive elements of the processor during idle phases.  
This establishes a direct correspondence between ancillary qubits, crossed elements, and energy-mediated quantum actuators, reinforcing their role as practical components for implementing programmable interactions in globally controlled quantum processors.

\section{Applications in superconducting global quantum computing}
\label{sec:applications}

In this section, we discuss how the integration of quantum actuators enables new functionalities in globally driven superconducting quantum processors.
In particular, we show that conceptual step~\autoref{conceptual2} naturally leads to programmable architectures in which different regions of the processor can be dynamically activated, frozen, or interconnected using only global controls.

\subsection{Programmability}

First, we now exploit conceptual step~\autoref{conceptual2} to introduce a form of {\it spatial programmability} in globally driven architectures. The key idea is to partition a large-scale processor into regions whose computational activity is conditionally enabled or suppressed by quantum actuators.

\begin{figure}[t]
    \centering
    \includegraphics[width=1.0\columnwidth]{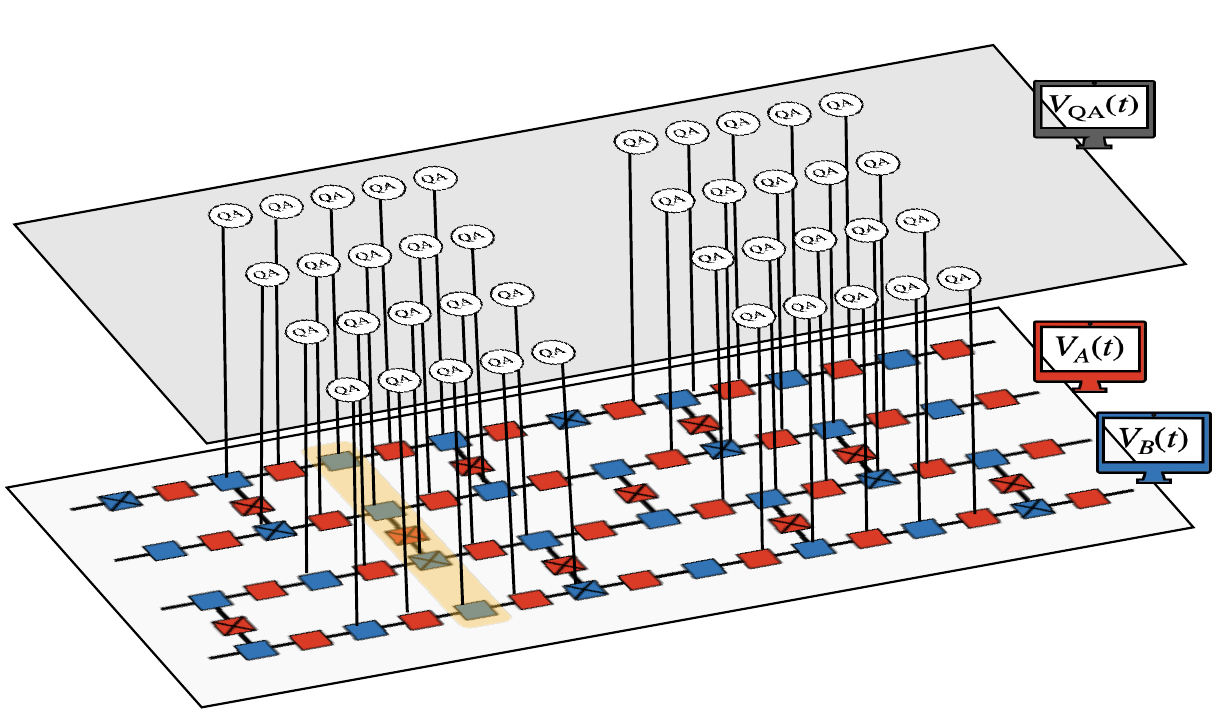}
    \caption{Programmability via globally driven quantum actuators in a globally driven quantum processor.
    A layer of auxiliary quantum actuators (QAs), shown schematically above the physical qubits, is globally controlled and selectively maintained in the excited state.
    Actuator-assisted blockade dynamically freezes the underlying regions of the processor, preventing them from participating in the computation. By locally disabling selected QA regions, active pathways (bridges) are created that allow coherent transport of the ICC across otherwise inactive areas. 
    }

    \label{fig:program}
\end{figure}

Consider, for concreteness, a ladder-like architecture as reviewed in Sec.~\ref{sec:global} and illustrated in Fig.~\ref{fig:ladder-cb}(a).
We imagine that, in selected regions of the ladder, each physical qubit is coupled to an auxiliary QA.
All actuators belonging to the same region are addressed by an additional global control line and are initialized and maintained in their excited state, as schematically shown in Fig.~\ref{fig:program}.

Due to the blockade-type interaction between a QA and its neighboring physical qubit, keeping the QA excited dynamically suppresses the effective action of the global driving fields on that qubit.
As a result, the corresponding region of the processor becomes {\it frozen}: physical qubits remain pinned to their instantaneous state and do not participate in the computational dynamics.
Importantly, this freezing mechanism does not require local addressing and is fully compatible with global control.

Conversely, in regions where the actuators are discharged or decoupled, the physical qubits regain their full dynamical response to the global drives and support coherent gate operations.
In this way, quantum actuators act as programmable energetic masks that define active and inactive zones of the processor.

This mechanism enables several useful operations.
First, it allows the selective confinement of the ICC to a desired subregion of the ladder.
Second, one can engineer {\it bridges} between distant active regions by locally disabling the actuators along a prescribed path.
In this configuration, the ICC can be coherently transported across otherwise frozen areas, effectively routing quantum information from one computational zone to another without modifying the underlying hardware (see Fig.~\ref{fig:program}).

The resulting programmability is purely dynamical and energy-based: regions are activated or deactivated by controlling the excitation state of the associated quantum actuators.
No additional local control lines are required, and the number of independent drive sources remains minimal.
This approach therefore addresses one of the main bottlenecks of scalable superconducting architectures, namely the wiring and control overhead associated with fine-grained addressability.

\subsection{Modularity}

Beyond spatial programmability within a single processor, quantum actuators enable a modular approach to globally driven quantum computation.
Here, modularity refers to the ability to dynamically connect and disconnect distinct computational units—such as independent conveyor-belt (CB) processors—using purely energetic and globally controlled mechanisms.

\begin{figure}[t]
    \centering
    \begin{overpic}[width=1.0\linewidth]{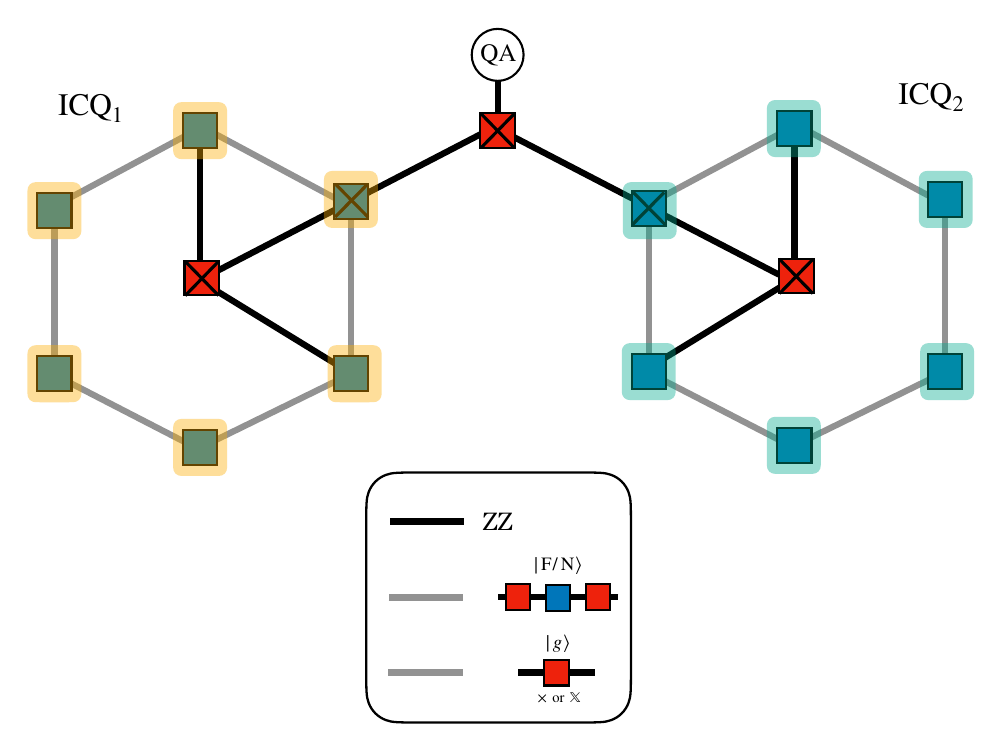}
    \put(40, 15.5){\cite{cioni2024conveyorbelt}}
    \put(40, 8.5){\cite{menta2025building}}
    \end{overpic}
    \caption{Modular interconnection of globally driven conveyor-belt processors via quantum actuators.
    Two independent conveyor-belt architectures are linked by a bridge region containing an auxiliary quantum actuator (QA). When the QA is maintained in its excited state, blockade-type interactions dynamically isolate the two processors. Switching the QA to its ground state activates the bridge, enabling coherent two-qubit operations between boundary computational qubits.
    A SWAP operation allows quantum information to be transferred between modules.}
    \label{fig:mod}
\end{figure}

Consider two conveyor-belt architectures, each encoding its own set of computational qubits, ${\rm ICQ}_{i}$
As schematically illustrated in Fig.~\ref{fig:mod}, the two CB modules are connected by a short bridge region containing an auxiliary quantum battery.
When the quantum battery is maintained in its excited state, blockade-type interactions suppress the effective dynamics on the bridge, and the two conveyor belts evolve as isolated quantum processors.

To enable inter-module communication, the quantum battery is switched to its ground state, lifting the blockade and activating the bridge region.
During this time window, appropriately designed global control sequences allow the implementation of coherent two-qubit gates between boundary computational qubits of the two CBs.
Let $Q_L$ and $Q_R$ denote the boundary qubits of the left and right conveyor belts, respectively.

A SWAP operation between $Q_L$ and $Q_R$ is implemented using the standard gate decomposition
\begin{equation}
\hat{U}_{\rm SWAP}^{(Q_L,Q_R)}
=
\hat{U}_{\rm CNOT}^{(Q_L \rightarrow Q_R)}
\hat{U}_{\rm CNOT}^{(Q_R \rightarrow Q_L)}
\hat{U}_{\rm CNOT}^{(Q_L \rightarrow Q_R)} \otimes \ket{g}_{\rm QA}\bra{g}\;.
\end{equation}
Each CNOT gate is natively realized within the globally driven architecture through the crossed-qubit and dynamical blockade mechanisms discussed in Sec.~\ref{sec:batteries}.

Crucially, each conveyor-belt module is driven by its own set of global control fields.
This allows one to selectively activate control sequences on one module while keeping the other frozen by re-exciting its associated quantum actuators.
As a result, the three CNOT operations composing the SWAP can be executed sequentially without violating the global-control constraints.

Once the SWAP operation is completed, the quantum battery is re-excited, restoring the blockade and dynamically decoupling the two conveyor-belt processors.
In this way, quantum information is coherently transferred between modules while preserving modular isolation outside the communication window.

This actuator-controlled coupling mechanism provides a minimal primitive for modular quantum computation.
It enables the construction of large-scale quantum processors composed of standardized globally driven units that can be dynamically interconnected for information transfer, initialization, or readout, and subsequently isolated again—without introducing additional local control lines or exacerbating the wiring overhead.

\section{Discussion}\label{sec:discussion}

In this work, we introduced quantum actuators as structural elements for globally controlled quantum computation.  
By reinterpreting ancillary and crossed qubits as mediators of interaction rather than as computational resources, we showed how selective gate activation, spatial programmability, and modular interconnection can be achieved under purely global control.

Our approach is grounded in two conceptual steps.  
First, we demonstrated that crossed elements naturally function as quantum actuators, enabling native multi-qubit gate operations through dynamical blockade mechanisms.  
Second, we showed that auxiliary actuators can be used to dynamically freeze or activate selected regions of a processor, providing programmable routing and modular connectivity without introducing local control lines.

Overall, our results position quantum actuators as architectural gadget for globally controlled quantum computation.  
They provide a mechanism for embedding controllable interaction layers within globally driven processors, extending their flexibility while preserving their defining features.  
At the same time, their energetic interpretation establishes a connection with concepts from quantum thermodynamics, offering a concrete setting in which such ideas acquire a direct operational role.

\acknowledgments
The authors are indebted with F. Caravelli for comments on the manuscript. 

\bibliography{biblio}
\end{document}